# Contexts of Diffusion: Adoption of Research Synthesis in Social Work and Women's Studies

Laura Sheble and Annie T. Chen

School of Information & Library Science
University of North Carolina, Chapel Hill, NC, USA, 27599
sheble@live.unc.edu, atchen@email.unc.edu

**Abstract**: Texts reveal the subjects of interest in research fields, and the values, beliefs, and practices of researchers. In this study, texts are examined through bibliometric mapping and topic modeling to provide a bird's eye view of the social dynamics associated with the diffusion of research synthesis methods in the contexts of Social Work and Women's Studies. Research synthesis texts are especially revealing because the methods, which include meta-analysis and systematic review, are reliant on the availability of past research and data, sometimes idealized as objective, egalitarian approaches to research evaluation, fundamentally tied to past research practices, and performed with the goal informing future research and practice. This study highlights the co-influence of past and subsequent research within research fields; illustrates dynamics of the diffusion process; and provides insight into the cultural contexts of research in Social Work and Women's Studies. This study suggests the potential to further develop bibliometric mapping and topic modeling techniques to inform research problem selection and resource allocation.

## 1 Introduction

Research synthesis methods (RSM), which include systematic reviews and meta-analysis, are among the most important recent methodological innovations in science, especially in applied fields. As RSM have diffused across science, the methods have been subject to adaptation and interpretation in the context of research fields. Because RSM typically require researchers to systematically engage with past research and empirically assess the relevance and strength of evidence presented in findings to integrate across studies, examination of how research synthesis is practiced in disciplines can provide insights about the nature of research, and values and beliefs of researchers within fields.

This study provides an overview of the dynamic processes associated with the diffusion of research synthesis in Social Work and Women's Studies. To an extent, the fields appear similar: each has a high proportion of women scholars; is aligned with interests of marginalized segments of society; and identified with the social sciences [1-2]. However, bibliometric mapping and topic modeling illustrate that the two fields exist in very different cultural contexts, as reflected in different patterns of RSM use and non-use; engagement with the knowledge base; and prevalence of meta-research topics. The current study is part of a larger study designed to investigate how, and in what ways adoption of research synthesis methods reflects and is shaped by the cultural contexts of science fields; and the application of bibliometric mapping and text analysis to reveal field-level research dynamics.

### 1.1 Research Contexts and Research Synthesis

Research synthesis involves systematic, empirical integration of science knowledge across research studies. Methods of research synthesis require researchers to work within the framework of their own research project; and negotiate the methods, data, and reporting structures of the primary studies that comprise the data for the synthesis study. Adoption of research synthesis methods within a field is facilitated by common and relatively stable approaches to research design, methods, measures, and problem selection; and belief structures that associate value with accumulations of research-based knowledge [3]. Information resources that enable access to data and research literature, and organizations that support synthetic research methods contribute to adoption of the methods by facilitating access to data for synthetic studies [4-5], and contributing to visibility of syntheses and requisite tools and techniques to perform such studies. In fields associated with professional practice and/or policy, engagement with the evidence based practice (EBP) movement provides impetus to use research synthesis methods, which are an integral component of EBP.

## 2 Data, Analysis, and Visualization

Data for this study was extracted from the Thomson Science *Science* and *Social Science Citation Indexes* (*S/SCI*). To estimate the extent to which fields engage with RSM the S/SCI was searched with keyword and citing reference queries [6]. "Diverse" forms of research synthesis methods, those designed for synthesis of qualitative or interpretive research, were represented by a subset of these queries. A keyword search, refined with iterative searching and scanning, was used to estimate the prevalence of engagement with evidence-based practice. The "Social Work" and "Women's Studies" Web of Science categories were used to delimit research fields. Identified publications were rated "directly," "indirectly," or "not apparently" related to research synthesis methods based on titles, author keywords, and abstracts. Only directly or indirectly related items are included in bibliometric maps and topic models.

### 2.1 Bibliometric Mapping

Publications referenced by Social Work and Women's Studies RSM papers were overlaid on a global map of science to identify the knowledge base of each field. Cosine-normalized citation patterns across science reported in the 2010 *Journal Citation Reports (JCR)*, aggregated by Web of Science categories, are the basis of the global science network [7]. Citation patterns represent cognitive or socio-cognitive similarity between science fields, for which the categories are considered a proxy. The base map was visualized in Pajek [8] and overlaid with counts of references in Social Work and Women's Studies publications. Nodes represent fields, and are sized in proportion to the number of references observed. Shannon evenness and Rao-Stirling diversity [9-10] describe the distribution of references across science fields. Shannon evenness is a ratio of Shannon entropy, which measures abundance and evenness of entities across categories, to the maximum entropy possible. Rao-Stirling diversity accounts for distribution across fields, and the degree to which fields differ. Difference between fields is estimated with the citation matrix used to construct the global map of science.

### 2.2 Topic Modeling

Topic models were developed to summarize RSM-related publications using a variational Bayes implementation of Latent Dirichlet Allocation (LDA) [11-13]. Publications were represented by word co-occurrences in titles, author keywords, and abstracts using a 'bag of words' approach. Text preprocessing included removing labels from structured abstracts, applying the Porter stemmer, and identifying stop words and frequently occurring words. The number of topics selected was informed by perplexity scores for 5 to 30 topics, such that local perplexity minima were preferred. Topic labels consist of word stems most frequently associated with each topic. The document-topic matrix, which describes topic distributions across documents, was visualized in Gephi [14] as a bimodal network. Network partitions were identified by the Louvain algorithm [15] and are represented by color. Topic nodes were sized in proportion to the sum of associated document proportions, and edge thresholds were applied to reduce visual complexity. Overall, this approach can be described as a quantitatively guided qualitative overview of publication content.

## 3 Contexts: RSM in Social Work and Women's Studies

The extent to which researchers engage with past research, evidence based practice, and diverse forms of research synthesis methods (i.e., forms amenable to qualitative and/or interpretive research) likely influence RSM diffusion within fields. Social Work and Women's Studies differ markedly with respect to these factors (Table 1). Evidence-based practice has been important in the field of Social Work, but only marginally so in Women's Studies. About 1.5% of all Social Work publications that appeared in the twenty years after EBP was popularized in medicine [16] are associated with EBP. In Women's Studies, this figure is roughly one-tenth of that observed in Social Work. Similarly, based on S/SCI document type classifications, Social Work researchers are about ten times more likely to publish reviews in the journal literature. Between 1976 and 2011, reviews comprised 1.33% of all Social Work publications and 0.12% of Women's Studies publications. Social Work has engaged with diverse forms of research synthesis about three times as often as Women's Studies. These results suggest that Social Work is a better fit for RSM.

**Table 1.** Relative prevalence of characteristics related to research synthesis methods

|  | *Social Work* | *Women's Studies* |
|---|---|---|
| Evidence based practice (1992 - ) | 1.49% | 0.15% |
| Reviews (S/SCI document type) | 1.33% | 0.12% |
| Year of first RSM publication | 1977 | 1985 |
| Diverse RS methods / all RSM | 6.73% | 2.33% |

Approximately two-thirds of all Social Work and one-third of all Women's Studies publications identified via the search protocol were judged as directly or indirectly related to research synthesis (Table 2). These relatively low (Social Work) and low (Women's Studies) proportions were unique to the social science fields selected from the larger study for more in-depth analysis. Review of results suggests that the S/SCI "KeywordsPlus" field, which includes words extracted from titles of cited references, contributed substantially to publications retrieved but apparently not related to research synthesis. Adherence to publishing practices and guidelines that proscribe indicating a paper reports a research synthesis in the title likely increases the prevalence of "meta-analysis" and similar terms in the KeywordsPlus field.

**Table 2.** Extent of relationship between retrieved publications and RSM

| Relationship | Social Work Count | Percent | Women's Studies Count | Percent |
|---|---|---|---|---|
| Direct | 284 | 57.96 | 95 | 31.77 |
| Indirect | 34 | 6.94 | 16 | 5.35 |
| Direct or Indirect | 318 | 64.9 | 111 | 37.12 |
| No apparent | 172 | 35.1 | 188 | 62.88 |
| Total | 490 |  | 299 |  |

### 3.1 Knowledge Base Interactions

Social Work and Women's Studies differ in the extent and diversity of references to the knowledge base. Though Social Work references a greater number of fields, most references are concentrated within Social Work and cognate fields. Lower Rao-Stirling diversity (Table 3) and the proximity of larger nodes to Social Work (Fig. 1, yellow node with red ring) reflect these differences. Women's Studies references are more evenly distributed across fields, as indicated by higher Shannon evenness and nodes that are more similar in size. Women's Studies seldom references work published in Women's Studies journals (Fig. 1, yellow node with red ring). The juxtaposition of a concentration of RSM publications within a few Women's Studies journals with broad referencing patterns suggests Women's Studies scholars engage with content of other fields through RSM. The pattern echoes observations that Women's Studies scholars tend to have dual allegiances: to Women's Studies and another field; and the description of feminist scholarship as one that "simultaneously challenges and is shaped by disciplinary inquiry" [17, p. 121]. Social Work, in contrast, mobilizes research in Social Work and cognitively similar fields for the benefit of the field broadly.

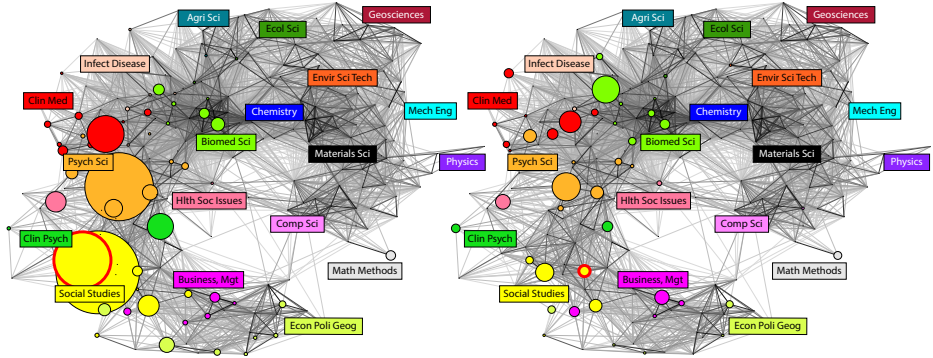

**Fig. 1.** Fields referenced by Social Work (left) and Women's Studies (right) RSM publications

**Table 3.** Diversity of fields referenced

| Measure | Social Work | Women's Studies |
|---|---|---|
| Fields referenced | 111 | 88 |
| Shannon evenness | 0.6361 | 0.7317 |
| Rao-Stirling diversity | 0.6299 | 0.7256 |

### 3.2 Diffusion Dynamics and Topics Associated with RSM

The temporal distribution of research synthesis publications in each field reveals a substantial lag between introduction and broader engagement with RSM. Across science broadly, the first research synthesis publications appeared in the early to mid 1970s [4,18], shortly before the first in Social Work (1977) and Women's Studies (1985). In Social Work, sustained engagement with the methods did not occur until the late 1990s, about twenty years after the first research synthesis publication in the field (Fig. 2). In Women's Studies, engagement with the methods has remained modest, but became more prevalent in 2003, and expanded again in 2009.

Topical distributions of research synthesis publications suggest broad engagement with research synthesis across Social Work; but selective and unevenly distributed engagement in Women's Studies. While each field is associated with 38 titles in the *JCR*, 36 of the Social Work titles, but only 14 of the Women's Studies titles include RSM publications. Concentration within journals is similarly uneven: 4 journals (10.5%) contain 81% of the RSM related publications in Women's Studies; and 17 journals (44.7%) contain 81% in Social Work.

Evidence-based practice, methodologies, and intervention research are salient issues in Social Work. In Fig. 2, these issues are represented by the topics "practice evidence-bas", "meta-analysis design result method", "effect size meta-analysis", "systematic search database", and "intervent treatment effect". The prevalence of top-ics associated with practice and EBP reflects the practice orientation of Social Work research [19]. EBP and RSM are central to discussions about research-practice divides; and have been cast in opposition to traditionally prevalent research approaches such as qualitative case studies [20], which are difficult to systematically synthesize. Since the late 1990s, interest in intervention research, which is prototypically amenable to synthesis methods, has emerged. Preventative programs ("program prevent") and client services ("service inform what client process"), which are linked to EBP, are also focal interests.

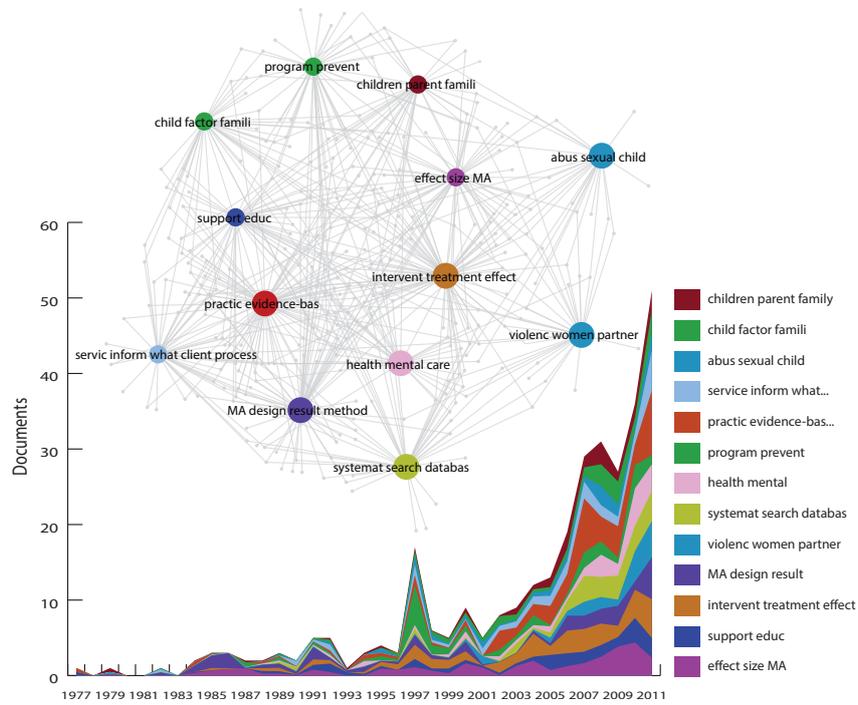

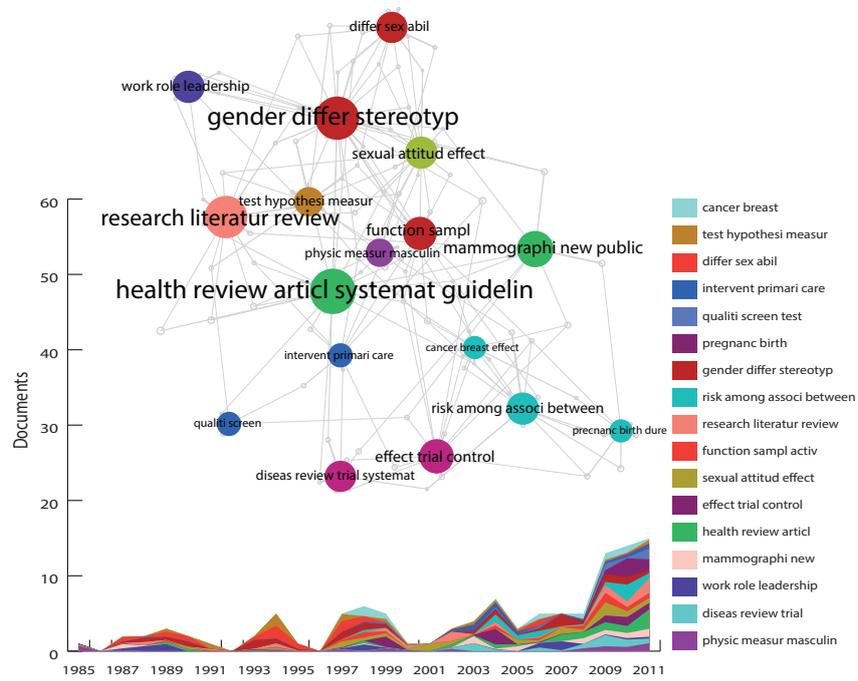

**Fig. 2.** Topics of Social Work (top) and Women's Studies (bottom) RSM publications

The limited range of topics and uneven distributions of content across topics provide evidence of selective engagement with RSM in Women's Studies. RSM publications are centered on research related to gender differences in psychology and related fields; women's health; and methodological issues (Fig. 2). Psychology topics include work and leadership ("work role leadership"), gender differences and stereotypes, and abilities linked with sex such as spatial navigation ("differ sex abil"). Women's health issues include breast cancer, mammography, reproductive issues, and conditions not specific to females. Topics that link health-oriented and psychological gender-oriented research include those focused on literature and synthesis ("research literature review", "health review articl systemat guidelin"), and methods (e.g., "function sampl"), though topic locations reveal emphases on different methods topics in health versus psychology fields. For example, "test hypothesi measure" is embedded in psychology topics; and interventions and guidelines are associated with health topics.

### 3.3 Summary: RSM in Social Work and Women's Studies

Contrasts in the cultures and approaches to research in Social Work and Women's Studies are reflected in the topics associated with research synthesis and the dynamics of the RSM diffusion process. In Social Work, engagement with EBP and a tradition that prioritizes social work practice over research has fundamentally shaped engagement. While a practice orientation provides motivation to synthesize, limitations associated with past research have moderated the applicability of prevalent research synthesis methods, and likely impeded diffusion. In Women's Studies, research synthesis methods have been used primarily in subfields associated with disciplines in which RSM are common - psychology and the health sciences. An activist stance is exemplified by engagement with research in cognate fields to communicate the value of the alternative lens Women's Studies offers; and to comment on prior research not compatible with addressing interests of diverse populations.

### 3.4 Conclusion and Future Research

Topic modeling and bibliometric mapping, in combination with domain expertise, can be used to visualize and analyze relationships and processes at the level of research fields. The methods were mobilized in this study to contrast dynamics and relationships associated with diffusion of research synthesis methods. Future research should focus on use of the methods to contrast differences within fields, including to examine research that is discussed prospectively versus that which is actually performed; to analyze and inform distribution and use of resources; and to provide high-level summaries of research in fields over time. Challenges include development and visualization of interpretable and comparable topic maps, accessibility of texts used to represent research, and flexible yet specifiable approaches to defining research fields.